\documentclass[final,3p,times,sort&compress]{elsarticle}

\usepackage{mathrsfs}        
\usepackage{graphicx}        
\usepackage{amssymb}         
\usepackage{amsmath}         
\usepackage{enumerate}       
\usepackage[colorlinks,
            linkcolor=red,
            anchorcolor=blue,
            citecolor=green
            ]{hyperref}
\usepackage{caption}
\usepackage{amsthm}          

\newtheorem{theorem}{Theorem}
\newtheorem{remark}{Remark}
\newtheorem{example}{Example}

\DeclareCaptionLabelSeparator{dotmark}{. }
\journal{Journal of The Franklin Institute}

\begin{document}

\begin{frontmatter}

\phantomsection
\addcontentsline{toc}{chapter}{Generalization of the gradient method with fractional order gradient direction}
\title{Generalization of the gradient method with fractional order gradient direction}


\author{Yiheng Wei, Yu Kang, Weidi Yin, Yong Wang}
\address{Department of Automation, University of Science and Technology of China, Hefei, 230026, China}


\begin{abstract}
Fractional calculus is an efficient tool, which has the potential to improve the performance of gradient methods. However, when the first order gradient direction is generalized by fractional order gradient one, the corresponding algorithm converges to the fractional extreme point of the target function which is not equal to the real extreme point. This drawback critically hampers the application of this method. To solve such a convergence problem, the current paper analyzes the specific reasons and proposes three possible solutions. Considering the long memory characteristics of fractional derivative, short memory principle is a prior choice. Apart from the truncation of memory length, two new methods are developed to reach the convergence. The former is the truncation of the infinite series, and the latter is the modification of the constant fractional order. Finally, six illustrative examples are performed to illustrate the effectiveness and practicability of proposed methods.
\end{abstract}


\end{frontmatter}


\section{Introduction}\label{Section1}
Gradient method, as a classical tool in optimization theory, because of its intuitive principle, simple structure and easy implementation, has attracted much attention and developed rapidly since its inception. Now it prevalently used in many fields, such as automatic control \cite{Ren:2019JFI}, system identification \cite{Ge:2019JFI}, machine learning \cite{Lecun:2015Nature}, and image denoising \cite{Pu:2010TIP}. Besides, many effective algorithms are based on it or inspired by it \cite{Boyd:2004Book}. Nonetheless, the traditional gradient descent method has the characteristics of ``zigzag'' descent and the disadvantage of slow convergence near the extreme value, which seriously affects the performance of its derivative optimization algorithm, and further limits their application in practice. As a result, it is the goal that researchers and engineers are constantly striving for to solve this problem and improve the performance further.

Study shows that the introduction of fractional calculus can effectively improve the convergence speed and optimization performance of gradient descent method. It mainly lies in its special long memory characteristics \cite{Liu:2017SMCA,Yin:2019MSSP}. In 2015, we initially proposed a gradient method with fractional order update style, i.e., replacing the first order derivative by fractional order derivative, and adopted it in parameter estimation \cite{Wei:2015JCD}. After considering the constraint condition, we designed the corresponding fractional projection gradient method \cite{Chen:2016CNSNS}. Afterwards, Tan proposed a class of fractional order LMS algorithm, in which the first order difference was replaced by fractional order difference \cite{Tan:2015SPL}. This can be regarded as the discrete case of  fractional order update style. Besides, other results on the application of this fractional order gradient method were reported in \cite{Cheng:2017ISA,Cui:2018ISA,Liang:2019OL}.
Also in 2015, a new fractional order gradient method was developed, where fractional order gradient direction was used as a substitute for the first order gradient direction \cite{Pu:2015TNNLS}. This new method manifests distinct properties. For example, its iterative search process can easily pass over the local extreme point. However, one cannot guarantee that the extreme point can be found using the method in \cite{Pu:2015TNNLS} even if the algorithm is indeed convergent. Additionally, it is difficult to calculate the needed fractional derivative online. These remarkable progress on fractional order gradient method not only reveals some interesting properties, but also gives practical inspiration for future research.

Actually, an available solution to overcome this shortcoming partially was implied in \cite{Zahoor:2009EJSR}, where the weighted sum of integer order derivative and fractional order derivative was adopted. Despite some minor errors when using chain rule of fractional derivative, the developed method can reduce the steady state error to a certain degree and has been successfully applied many scenarios, such as time series prediction \cite{Shoaib:2014CPB}, active noise control \cite{Aslam:2015SP}, speech enhancement \cite{Geravanchizadeh:2014IJEEE}, channel equalization \cite{Shah:2017NODY}, recommender systems \cite{Khan:2018EM}, RBF neural network \cite{Khan:2018CSSPa}, RNN neural network \cite{Khan:2018CSSPb}, etc. To avoid the error, an alternative solution with approximation fractional derivative was proposed by us \cite{Chen:2017AMC}. It was shown that this method resulted a better convergence performance than that in \cite{Zahoor:2009EJSR}. Hereinafter, this method was widely used in many applications, such as system identification \cite{Cui:2017NODY}, LMS algorithm \cite{Cheng:2017SP}, BP neural networks \cite{Wang:2017NN}, and CNN \cite{Sheng:2019Neu}.

It is worth pointing out that fractional gradient method has been used successfully, but the related research is still in its infancy and deserves further investigation. The immediate problem is the convergence problem. \cite{Du:2016CCC} has found that the fractional extreme value is not equal to the real extreme value of target function, which would make fractional gradient method loss practicability. However, the main reason for nonconvergence is unclear. On one hand, a systematic interpretation for this question is still desirable. On the other hand, more effective solutions for realizing convergence are needed. What is more, it is difficult to calculate the fractional order gradient of the target function in real time. Bearing this in mind, the objective of this paper are: i) investigating the specific reason for the nonconvergence of fractional order gradient method; ii) analyzing the relationship between the fractional extreme point and the nonlocality; iii) designing the solutions to solve the convergence problem; and iv) providing online computation method for fractional derivative.

The remainder of the paper is organized as follows. Section \ref{Section 2} is devoted to math preparation and problem formulation. Solutions to the convergence problem of fractional order gradient method are introduced in Section \ref{Section 3}. Section \ref{Section 4} shows some numerical examples to verify the proposed methods. At last, conclusion is presented in Section \ref{Section 5}.

\section{Preliminaries}\label{Section 2}
This section presents a brief introduction to the mathematical background of fractional calculus and the fatal flaw of fractional order gradient method.
\subsection{Fractional calculus}

The Riemann--Liouville derivative and Caputo derivative of function $f(\cdot)$, are expressed as \cite{Podlubny:1999Book}
\begin{equation}\label{Eq1}
{\textstyle{}_{{c}}^{{\rm{RL}}}\mathscr{D}_x^\alpha f(x) = \frac{1}{{\mathrm{\Gamma} (n - \alpha )}}\frac{{{\mathrm{d}^n}}}{{\mathrm{d}{x^n}}}\int_{{c}}^x {\frac{{f(\tau )}}{{{{(x - \tau )}^{ \alpha  - n+1}}}}} \mathrm{d}\tau,}
\end{equation}
\begin{equation}\label{Eq2}
{\textstyle{}_{{c}}^{{\rm{C}}}\mathscr{D}_x^\alpha f(x) = \frac{1}{{\mathrm{\Gamma} (n - \alpha)}}\int_{{c}}^x {\frac{{{f^{(n)}}(\tau )}}{{{{(x - \tau )}^{\alpha  - n + 1}}}}\mathrm{d}\tau },}
\end{equation}
respectively, where $n-1< \alpha < n$, $n  \in \mathbb{Z}_+$, $c$ is the lower integral terminal, $\Gamma (\alpha ) = \textstyle{\int_0^{ + \infty } {{{\rm e}^{ - t}}{t^{\alpha  - 1}}\mathrm{d}t}}$ is the Gamma function. Notably, the fractional derivatives in (\ref{Eq1}) and (\ref{Eq2}) are actually special integrals and they manifest long memory characteristic of the function $f(x)$, sometimes called nonlocality.

If function $f(x)$ can be expanded as a Taylor series, the fractional derivatives can be rewritten as \cite{Wei:2019ArXiv}
\begin{equation}\label{Eq3}
{\textstyle{}_{c}^{{\rm{RL}}}\mathscr{D}_x^\alpha f(x) = \sum\nolimits_{i = 0}^{ + \infty } {\big( {\begin{smallmatrix}
\alpha\\
i
\end{smallmatrix}} \big)}\frac{{{f^{(i)}}(x)}}{{\Gamma (i + 1 - \alpha )}}{{(x - c)}^{i - \alpha }},}
\end{equation}
\begin{equation}\label{Eq4}
{\textstyle{}_{c}^{{\rm{C}}}\mathscr{D}_x^\alpha f(x) = \sum\nolimits_{i = n}^{+\infty}{\big( {\begin{smallmatrix}
\alpha-n\\
i-n
\end{smallmatrix}} \big)}{\frac{{{f^{(i)}}({x})}}{{\mathrm{\Gamma} (i + 1 - \alpha )}}} {(x - c)^{i - \alpha }},}
\end{equation}
where $\big( {\begin{smallmatrix}
p\\
q
\end{smallmatrix}} \big)=\frac{{\Gamma (p  + 1)}}{{\Gamma (q + 1)\Gamma (p  - q + 1)}}$, $p\in\mathbb{R}$, $q\in\mathbb{N}$ is the generalized binomial coefficient. From the two formulas, it can be directly concluded that the fractional derivative, no matter for Riemann--Liouville definition or Caputo definition, consists of various integer order derivatives at $x$.

In general, the fractional derivative can be regarded as the natural generalization of the conventional derivative. However, it is worth mentioning that the fractional derivative has the special long memory characteristic, which will play a pivotal role in fractional order gradient method.

\subsection{Problem statement}
The well-known gradient descent method is a first order iterative optimization algorithm for finding the minimum of a function. To this end, one typically takes steps proportional to the negative of the gradient (or approximate gradient) of the function at the current point. For example, $x_k$ is updated by the following law
\begin{equation}\label{Eq5}
{\textstyle{x_{k + 1}} = {x_k} - \mu\nabla f({x_k}),}
\end{equation}
where $x_k$ is the current position, $x_{k+1}$ is the next position, $\mu$ is the learning rate and $\nabla f({x_k})$ is the first order gradient at $x=x_k$, i.e., $\frac{{\rm{d}}}{{{\rm{d}}x}}f\left( x \right)|_{x=x_k}$. When the classical gradient is replaced by the fractional one, it follows
\begin{equation}\label{Eq6}
{\textstyle{x_{k + 1}} = {x_k} - \mu\nabla^\alpha f({x_k}).}
\end{equation}
Due to the uniformity of fractional derivative, namely, $\mathop {\lim }\limits_{\alpha  \to 1} {}_c^{\rm RL}{\mathscr D}_x^\alpha f\left( x \right) = \mathop {\lim }\limits_{\alpha  \to 1} {}_c^{\rm C}{\mathscr D}_x^\alpha f\left( x \right) = \frac{{\rm{d}}}{{{\rm{d}}x}}f\left( x \right)$, $0<\alpha\le1$, (\ref{Eq6}) degenerates into (\ref{Eq5}) exactly when $\alpha=1$.

For $f(x)=(x-5)^2$, its real extreme point is $x^*=5$. Its first order derivative is $f^{(1)}(x) = 2x-10$,
while for $\alpha\in(0,1)$, its fractional order derivative satisfies
\begin{equation}\label{Eq7}
{\textstyle
\begin{array}{rl}
{}_{c}^{{\rm{RL}}}\mathscr{D}_x^\alpha f(x)=\frac{2}{{\Gamma (3 - \alpha )}}{(x - {c})^{2 - \alpha }}+\frac{{2(c - 5)}}{{\Gamma (2 - \alpha )}}{(x - {c})^{1 - \alpha }} +\frac{{(c - 5)}^2}{{\Gamma (1 - \alpha )}}{(x - {c})^{- \alpha }},
\end{array}}
\end{equation}
\begin{equation}\label{Eq8}
{\textstyle {}_{c}^{{\rm{C}}}\mathscr{D}_x^\alpha f(x)=\frac{2}{{\Gamma (3 - \alpha )}}{(x - {c})^{2 - \alpha }}+\frac{{2(c - 5)}}{{\Gamma (2 - \alpha )}}{(x - {c})^{1 - \alpha }} .}
\end{equation}

\begin{figure}[htbp]
\centering
\includegraphics[width=0.5\textwidth]{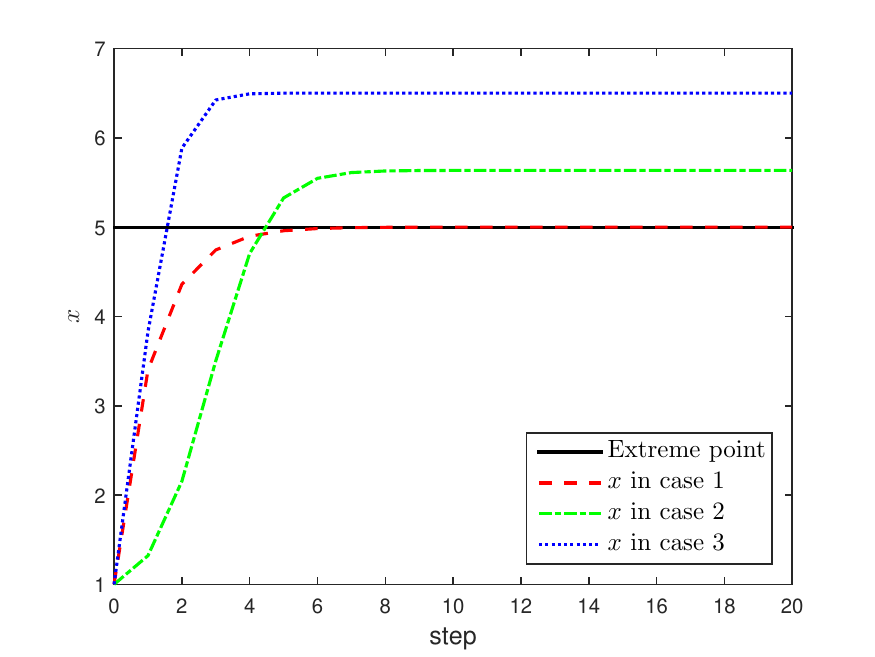}
\centering
\caption{Search process of gradient descent method.}\label{Fig 1}
\end{figure}

Consider the following three cases
\[
{\textstyle
\left\{\begin{array}{l}
\textrm{case 1: using (\ref{Eq6}) with first order derivative;}\\
\textrm{case 2: using (\ref{Eq6}) with Riemann--Liouville derivative;}\\
\textrm{case 3: using (\ref{Eq6}) with Caputo derivative,}
\end{array}\right.}
\]
and set $c=0$, $x_0=1$, $\mu=0.3$ and $\alpha=0.7$. On this basis, the simulation results are shown in Fig. \ref{Fig 1}.

It is clearly observed that all the three cases can realize the convergence within $20$ steps, while only case 1 is able to converge to the exact extreme point. If the algorithm is convergent for case 2, the convergent point satisfies ${}_{c}^{{\rm{RL}}}\mathscr{D}_x^\alpha f(x)=0$. With the exception of $x=c$, one obtains
\begin{equation}\label{Eq9}
{\textstyle \begin{array}{rl}
x =&\hspace{-6pt} c - \left( {c - 5} \right)\frac{{\left( {2 - \alpha } \right) \pm \sqrt {{{\left( {2 - \alpha } \right)}^2} - 2\left( {2 - \alpha } \right)\left( {1 - \alpha } \right)} }}{2}\\
 =&\hspace{-6pt} 3.25 \pm 2.5\sqrt {0.91} \\
 \approx&\hspace{-6pt} 5.6348 \textrm{~or~} 0.8652.
\end{array} }
\end{equation}
which matches the green dot dash line well. Moreover, if the initial search point $x_0$ is selected appropriately, then another extreme point $0.8652$ can be reached.

Similarly, when case 3 is convergent, the final value can be calculated by ${}_{c}^{{\rm{C}}}\mathscr{D}_x^\alpha f(x)=0$, which leads to
\begin{equation}\label{Eq10}
{\textstyle x = c - \left( {c - 5} \right)\left( {2 - \alpha } \right) = 6.5.}
\end{equation}
This just coincides with the blue dotted line.

From the previous discussion and additional simulation, it can be observed that fractional order gradient method does not always converge. Once it is convergent, the resulting extreme point is called the fractional extreme point. In general, such a fractional extreme point is different from the real extreme point. Because of the nonlocality, the fractional derivative of $f(x)$ depends on the variable $x$ and is also connected with the gradient order $\alpha$, the lower integral terminal $c$ and initial search point $x_0$. Thus, it is not easy to ensure ${}_{c}^{{\rm{RL}}}\mathscr{D}_x^\alpha f(x)=0$ and ${}_{c}^{{\rm{C}}}\mathscr{D}_x^\alpha f(x)=0$ at the real extreme point $x^*$. Now, the main reason can be summarized as follows.
\begin{enumerate}[i)]
  \item There are different definitions for fractional derivatives.
  \item Fractional derivatives have nonlocality while the classical derivative has locality
  \item The gradient order $\alpha$ may affect the convergence result.
  \item The lower integral terminal $c$ may affect the convergence result.
  \item The initial search point $x_0$ may affect the convergence result.
\end{enumerate}

Without loss of generality, consider $f(x)=a(x-x^*)^2+f_m$, $a>0$, $x^*\in\mathbb{R}$, $f_m\in\mathbb{R}$ and then similar conclusions can be drawn. If $\alpha\in(0,1)$ and $c\neq x^*$, then case 2 may not converge. Even if it is convergent, it will never converge to $x^*$. Besides, case 3 will never converge to $x^*$ either. This kind of convergence problem inevitably will, to some extent, cause performance deterioration when using fractional order gradient method in some practical applications. Therefore, it is necessary to deal with the convergence problem mentioned above.

\section{Main Results}\label{Section 3}
To solve the convergence problem, three viable solutions are proposed by fully considering the long memory characteristic of fractional derivative. Additionally, the convergence and implementation are described in detail.
\subsection{Iterate the lower integral terminal}
From the discussion in the previous section, it can be found that the fractional order gradient method cannot converge to the real extreme point as the order extends to the non-integer case. To substantially study the short memory characteristic of classical derivative and modify the long memory characteristic of fractional derivative, an intuitive idea is to replace the constant lower integral terminal $c$ with the varying lower integral terminal $x_{k-K}$. Then, fixed memory step appears and the resulting method follows immediately
\begin{equation}\label{Eq11}
{\textstyle{x_{k + 1}} = {x_k} - \mu { {{}_{{x_{k - K}}}^{\hspace{9pt}\rm{C}}{\mathscr D}_x^\alpha f\left( x \right)} |_{x = {x_k}}},}
\end{equation}
where $0<\alpha<1$, $\mu>0$ and $K\in\mathbb{Z}_+$.

This design is inspired by our previous work in \cite{Wei:2017FCAA,Chen:2017AMC} (see Example 4.3 of \cite{Wei:2017FCAA} and Theorem 2 of \cite{Chen:2017AMC}). On this basis, a theorem can be provided here.

\begin{theorem}\label{Theorem 1}
When the algorithm in (\ref{Eq11}) is convergent, it will converge to the real extreme point of $f\left( x \right)$.
\end{theorem}
\begin{proof}
Now, let us prove that $x_k$ converges to $x^*$ by contradiction. Suppose that $x_k$ converges to a point $X$ different from $x^*$ and $f^{(1)}(X)\neq0$, i.e., $\mathop {\lim }\limits_{k \to  + \infty } {x_k} = X$. Therefore, for any sufficient small positive scalar $\varepsilon$, there exists a sufficient large number $N\in\mathbb{N}$ such that $|x_k-X| < \varepsilon < |x^*-X|$ for any $k > N$.

By combining formulas (\ref{Eq4}) and (\ref{Eq11}), one has the following inequality
\begin{equation}\label{Eq12}
{\textstyle\begin{array}{rl}
\left| {{x_{k + 1}} - {x_k}} \right| =&\hspace{-6pt} \mu \big| {{{ {{}_{{x_{k - K}}}^{\hspace{9pt}\rm{C}}{\mathscr D}_x^\alpha f\left( x \right)} |}_{x = {x_k}}}} \big|\\
 =&\hspace{-6pt} \mu \big| {\sum\nolimits_{i = 1}^{{\rm{ + }}\infty } {{\big( {\begin{smallmatrix}
\alpha-1\\
i-1
\end{smallmatrix}} \big)}\frac{{{f^{\left( i \right)}}\left( {{x_k}} \right)}}{{\Gamma \left( {i + 1 - \alpha } \right)}}{{( {{x_k} - {x_{k - K}}} )}^{i - \alpha }}} } \big|\\
 \ge&\hspace{-6pt} \mu \sigma \sum\nolimits_{i = 0}^{{\rm{ + }}\infty } {{{\left| {{x_k} - {x_{k - K}}} \right|}^i}} {\left| {{x_k} - {x_{k - K}}} \right|^{1 - \alpha }}\\
 =&\hspace{-6pt} \mu \sigma \frac{{{{\left| {{x_k} - {x_{k - K}}} \right|}^{1 - \alpha }}}}{{1 - \left| {{x_k} - {x_{k - K}}} \right|}}\\
 \ge&\hspace{-6pt} d{\left| {{x_k} - {x_{k - K}}} \right|^{1 - \alpha }},
\end{array}}
\end{equation}
where $\sigma  = \mathop {\sup }\limits_{k > N,i \in \mathbb{N}} {\big( {\begin{smallmatrix}
\alpha-1\\
i
\end{smallmatrix}} \big)}\frac{{{f^{\left( {i + 1} \right)}}\left( {{x_k}} \right)}}{{\Gamma \left( {i + 2 - \alpha } \right)}}$ and $d = \frac{{\mu \sigma }}{{1 - \varepsilon }} $.

Actually, one can always find an $\varepsilon$ satisfying $d> \frac{2}{K^{1-\alpha}}{\varepsilon^{\alpha} }$. Hence, it follows
\begin{equation} \label{Eq13}
{\textstyle
\left| {{x_{k + 1}} - {x_k}} \right| > d\left| K\varepsilon\right|^{1-\alpha} > 2\varepsilon.}
\end{equation}

From the assumption on $|x_k-X| < \varepsilon < |x^*-X|$, it becomes
\begin{equation} \label{Eq14}
{\textstyle
\left| {{x_{k + 1}} - {x_k}} \right| \le \left| {{x_{k + 1}} - X} \right| +\left| {{x_{k}} - X} \right|  < 2\varepsilon,}
\end{equation}
which contradicts to the fact in (\ref{Eq13}). In other words, if the algorithm (\ref{Eq11}) converges to $X$, then $X$ must be equal to the real extreme point $x^*$. This completes the proof of Theorem \ref{Theorem 1}.
\end{proof}

The main idea of this method is called as short memory principle \cite{Wei:2017FCAA}. ${ {{}_{{x_{k - K}}}^{\hspace{9pt}\rm{C}}{\mathscr D}_x^\alpha f\left( x \right)} |_{x = {x_k}}}$ can be calculated with the help of (\ref{Eq4}). Then the corresponding fractional gradient method can be expressed as
\begin{equation} \label{Eq15}
{\textstyle{x_{k + 1}} = {x_k} - \mu \sum\nolimits_{i = 1}^{ + \infty } {{\big( {\begin{smallmatrix}
\alpha-1\\
i-1
\end{smallmatrix}} \big)}\frac{{{f^{(i)}}({x_{k}})}}{{\Gamma (i + 1 - \alpha )}}{{({x_k} - {x_{k - K}})}^{i - \alpha }}}.}
\end{equation}
Notably, if the algorithm (\ref{Eq15}) is convergent, it can converge to the real extreme point for any $\alpha>0$. The method in (\ref{Eq4}) could work effectively and efficiently if finite terms are enough to describe $f(\cdot)$ with a good precision. For the algorithm in (\ref{Eq11}), if the analytical form of ${ {{}_{{x_{k - K}}}^{\hspace{9pt}\rm{C}}{\mathscr D}_x^\alpha f\left( x \right)} |_{x = {x_k}}}$ can be obtained directly, there is no computational burden. For the algorithm in (\ref{Eq15}), it is generally impossible to use directly. If only finite terms are nonzero or finite terms approximation is enough, then the computational complexity is acceptable.

To facilitate the understanding, the proposed algorithm is briefly introduced in $\bf{Algorithm\ 1}$.
\begin{table}[!htbp]
\centering
\small
\renewcommand\arraystretch{1.5}
\begin{tabular*}{10.5cm}{lll}
\hline
\leftline{${\bf{Algorithm\ 1}}\quad \textrm{Fractional order gradient method with fixed memory step}.$}\\
\hline
${\bf{Input:}}\ x_0,x_1,\cdots,x_{K-1}$\\
${\bf{Output:}}\ x_N$\\
$\bf{Initialization:}$\\
$\quad\alpha,\mu,N,K:\ \rm{user\ defined\ value}$\\
${\bf{for}}\ k\ =\ K\ {\rm{to}}\ N-1\ {\bf{do}}$\\
$\quad h=\sum\nolimits_{i = 1}^{ + \infty } {{\big( {\begin{smallmatrix}
\alpha-1\\
i-1
\end{smallmatrix}} \big)}\frac{{{f^{(i )}}({x_{k}})}}{{\Gamma (i + 1 - \alpha )}}{{({x_k} - {x_{k - K}})}^{i  - \alpha }}}$\\
$\quad {x_{k + 1}} = {x_k} - \mu  h$\\
$\bf{end\ for}$\\
\hline
\end{tabular*}
\end{table}
\vspace{-3pt}

\subsection{Truncate higher order terms}
Recalling the classical gradient method, it can be obtained that for any positive $\mu$, if it is small enough, one has $f\left( {{x_k} - \mu } \right) \approx f\left( {{x_k}} \right) - \mu {f^{\left( 1 \right)}}\left( {{x_k}} \right)$.
Moreover, it becomes
\begin{equation}\label{Eq16}
{\textstyle
\begin{array}{rl}
f\left( {x_{k+1}} \right)=&\hspace{-6pt}f( {{x_k} - \mu {f^{\left( 1 \right)}}\left( {{x_k}} \right)} )\\
\approx&\hspace{-6pt} f\left( {{x_k}} \right) - \mu {[ {{f^{\left( 1 \right)}}\left( {{x_k}} \right)} ]^2}\\
\le&\hspace{-6pt} f\left( {{x_k}} \right).
\end{array}}
\end{equation}
When the dominant factor ${{f^{\left( 1 \right)}}\left( {{x_k}} \right)}=0$ emerges, the iteration completes, which confirms the fact that the first order derivative is equal to $0$ at the extreme point. From this point of view, only the relevant term is reserved and the other terms are omitted, resulting a new fractional gradient method
\begin{equation} \label{Eq17}
{\textstyle{x_{k + 1}} = {x_k} - \mu \frac{{{f^{(1)}}(x_k)}}{{\Gamma (2 - \alpha )}}{(x_k - {c})^{1 - \alpha }},}
\end{equation}
where $0<\alpha<1$, $x_0\neq c$ and $\mu>0$.

To avoid the appearance of a complex number, the update law in (\ref{Eq15}) can be rewritten as
\begin{equation} \label{Eq18}
{\textstyle{x_{k + 1}} = {x_k} - \mu \frac{{{f^{(1)}}(x_k)}}{{\Gamma (2 - \alpha )}}{|x_k - {c}|^{1 - \alpha }}.}
\end{equation}
To prevent the emergence of a denominator of $0$, i.e., $x_k=c$, a small nonnegative number $\epsilon$ is introduced to modify the update law further as
\begin{equation} \label{Eq19}
{\textstyle{x_{k + 1}} = {x_k} - \mu \frac{{{f^{(1)}}(x_k)}}{{\Gamma (2 - \alpha )}}{(|x_k - {c}|+\epsilon)^{1 - \alpha }}.}
\end{equation}

Similarly, one can determine whether the algorithm will converge to the real extreme point as Theorem \ref{Theorem 2}.
\begin{theorem}\label{Theorem 2}
When the algorithm in (\ref{Eq19}) is convergent, it will converge to the real extreme point of $f\left( x \right)$.
\end{theorem}
This theorem can be proved in the similar method like Theorem \ref{Theorem 1}. Here, a discussion can be provided from a different view. Algorithm (\ref{Eq19}) can be regarded as a varying learning rate case ${x_{k + 1}} = {x_k} - \mu (k)\nabla{f}({x_k})$ with $\mu (k) = \frac{1}{{\Gamma (2 - \alpha )}}{(|{x_k} - c| + \epsilon)^{1 - \alpha }}$. When $x_k$ tends to the extreme point, $\mu (k)$ tends to a constant. $\mathop {\lim }\limits_{k \to  + \infty } \nabla{f}({x_k}) = 0$ will emerge in the limiting case. Assuming that $x^*$ is the unique extreme point of ${{f}\left( \cdot\right)}$, then one has $\mathop {\lim }\limits_{k \to  + \infty } {x_k} = {x^*}$. Compared with the traditional gradient method, only $\mu$ is extended to $\mu(k)$, computational complexity will increase a little while it will bring computational burden.

Similarly, a brief description of the algorithm is shown in $\bf{Algorithm\ 2}$.
\begin{table}[!htbp]
\centering
\small
\renewcommand\arraystretch{1.5}
\begin{tabular*}{10.5cm}{lll}
\hline
\leftline{${\bf{Algorithm\ 2}}\quad \textrm{Fractional order gradient method with higher order truncation}.$}\\
\hline
${\bf{Input:}}\ x_0$\\
${\bf{Output:}}\ x_N$\\
$\bf{Initialization:}$\\
$\quad\alpha,\mu,N,c,\epsilon:\ \rm{user\ defined\ value}$\\
${\bf{for}}\ k\ =\ 0\ {\rm{to}}\ N-1\ {\bf{do}}$\\
$\quad h=\frac{{{f^{(1)}}(x_k)}}{{\Gamma (2 - \alpha )}}{(|x_k - {c}|+\epsilon)^{1 - \alpha }}$\\
$\quad {x_{k + 1}} = {x_k} - \mu h$\\
$\bf{end\ for}$\\
\hline
\end{tabular*}
\end{table}
\vspace{-3pt}

\subsection{Design variable fractional order}
It is well known that the traditional gradient method could converge to the exact extreme point. For this reason, adjusting the order $\alpha$ with $x$ is an alternative method. If the target function satisfies $f(x)\ge 0$ and has unique extreme point $x^*$, one can design the variable fractional order as follows
\begin{equation}\label{Eq20}
{\textstyle\alpha(x)  = \frac{1}{{1 + {\beta J(x)}}},}
\end{equation}
\begin{equation}\label{Eq21}
{\textstyle\alpha(x)  = \frac{2}{{1 + {{\rm e}^{\beta J(x)}}}},}
\end{equation}
\begin{equation}\label{Eq22}
{\textstyle\alpha(x)  = \frac{1}{{\rm cosh}{\beta J(x)}},}
\end{equation}
\begin{equation}\label{Eq23}
{\textstyle\alpha(x)  =1- \frac{2}{\pi}{\rm atan} \left( {\beta J(x)} \right),}
\end{equation}
\begin{equation}\label{Eq24}
{\textstyle\alpha(x) = 1 - \tanh \left( {\beta J(x)} \right),}
\end{equation}
where the loss function $J(x)=f(x)$ and the constant $\beta>0$. Besides, it is noticed that
\begin{equation}\label{Eq25}
{\textstyle 0 < \alpha \left( x \right) < 1,\beta  \ne 0,x \ne x^*,}
\end{equation}
\begin{equation}\label{Eq26}
{\textstyle \mathop {\lim }\limits_{\beta  \to 0} \alpha \left( x \right) = \mathop {\lim }\limits_{x \to x^*} \alpha \left( x \right) = 1.}
\end{equation}

\begin{figure}[!htbp]
\centering
\includegraphics[width=0.5\textwidth]{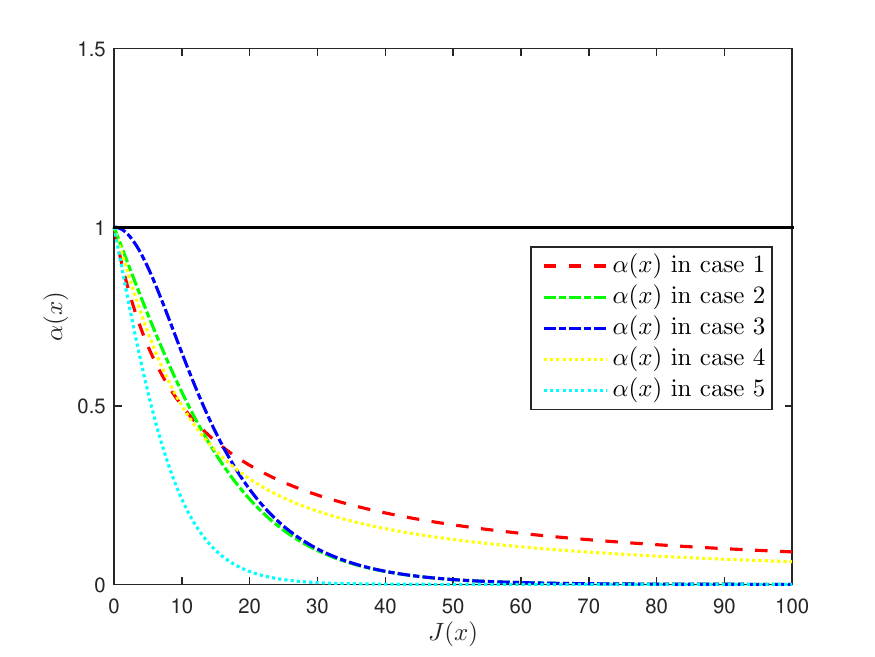}\includegraphics[width=0.5\textwidth]{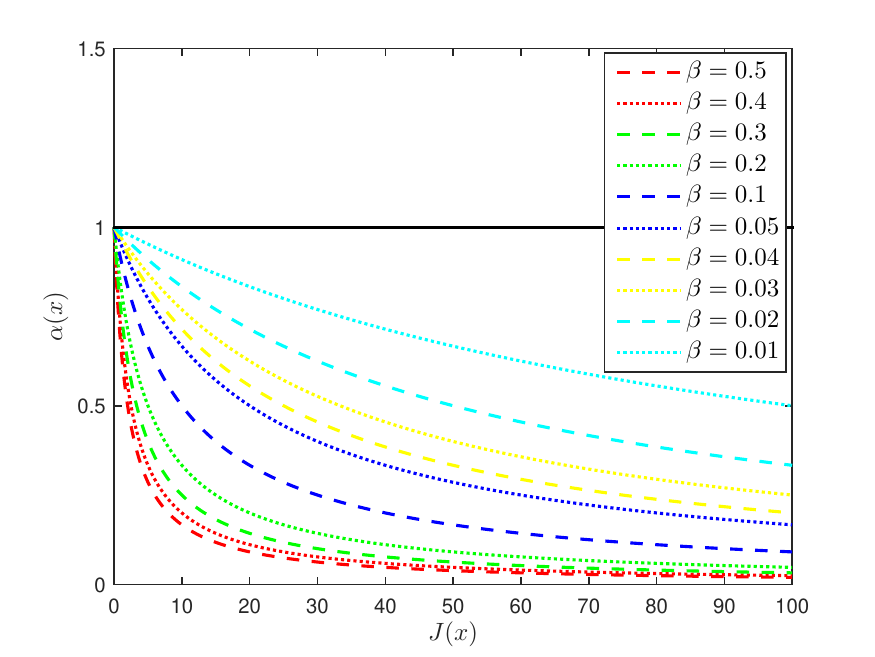}
\centering
\caption{The change of $\alpha(x)$ (left: three cases with $\beta=0.1$, right: case 1 with different $\beta$).}\label{Fig 2}
\end{figure}

To get an intuitive understanding of $\alpha(x)$, let us consider the following situations
\begin{equation}\label{Eq27}
{\textstyle
\left\{\begin{array}{l}
\textrm{case 1: using variable fractional order in (\ref{Eq20});}\\
\textrm{case 2: using variable fractional order in (\ref{Eq21});}\\
\textrm{case 3: using variable fractional order in (\ref{Eq22});}\\
\textrm{case 4: using variable fractional order in (\ref{Eq23});}\\
\textrm{case 5: using variable fractional order in (\ref{Eq24}),}
\end{array}\right.}
\end{equation}
and then the simulation results can be obtained in Fig. \ref{Fig 2}.

From Fig. \ref{Fig 2}, it can be concluded that when $J(x)\approx0$, one has $\alpha(x)\approx 1$. When $J(x)\gg0$, one has $\alpha(x)\approx 0$. Additionally, larger $\beta$ corresponds to quicker change as $J(x)$ is close to $0$.

With this design, $x$ is far from $x^*$ at the beginning of learning, and then $\alpha(x)\approx 1$, which results in a quick learning. As the learning proceeds, $x$ get close to $x^*$ gradually and then $\alpha(x)\approx1$, which leads to an accurate learning. In the end, i.e., $k\to +\infty$, $x \to x^*$ is expected.

However, the order $\alpha (x)$ is constructed with the assumption on $f(x)$. If the minimum value of $f(x)$ is nonzero or even negative, the designed orders in (\ref{Eq20})-(\ref{Eq24}) will no longer work. In this case, the function will be redefined
\begin{equation}\label{Eq28}
{\textstyle J\left( x \right) = {[{f^{(1)}}(x)]^{2}},}
\end{equation}
and thereby the designed orders are revived. At this point, the corresponding method can be expressed as
\begin{equation} \label{Eq29}
{\textstyle{x_{k + 1}} = {x_k} - \mu \sum\nolimits_{i = 1}^{ + \infty } {{\big( {\begin{smallmatrix}
\alpha(x)-1\\
i-1
\end{smallmatrix}} \big)}\frac{{{f^{(i)}}({x_{k}})}}{{\Gamma (i + 1 - \alpha(x) )}}{{({x_k} - c)}^{i - \alpha(x) }}},}
\end{equation}
where $x_0\neq c$ and $\mu>0$.

Likewise, with the specially designed $\alpha(x)$, the following theorem follows.
\begin{theorem}\label{Theorem 3}
When the algorithm in (\ref{Eq29}) is convergent, it will converge to the real extreme point of $f\left( x \right)$.
\end{theorem}

To maintain the conciseness of the representation, the proof is strategically omitted. With regard to the computational complexity, this variable fractional order method  performs a little worse than the fixed memory step method. When tracking differentiator or some other efficient numerical techniques are adopted to calculate ${ {{}_{{x_{k - K}}}^{\hspace{9pt}\rm{C}}{\mathscr D}_x^{\alpha(x)} f\left( x \right)} |_{x = {x_k}}}$, the calculation will not be a problem.

The description of the discussed algorithm is given in $\bf{Algorithm\ 3}$.
\begin{table}[!htbp]
\centering
\small
\renewcommand\arraystretch{1.5}
\begin{tabular*}{10.5cm}{lll}
\hline
\leftline{${\bf{Algorithm\ 3}}\quad \textrm{Fractional order gradient method with variable fractional order}.$}\\
\hline
${\bf{Input:}}\ x_0$\\
${\bf{Output:}}\ x_N$\\
$\bf{Initialization:}$\\
$\quad\mu,N,c,\beta:\ \rm{user\ defined\ value}$\\
${\bf{for}}\ k\ =\ 1\ {\rm{to}}\ N-1\ {\bf{do}}$\\
$\quad \alpha(x) $ in (\ref{Eq20}) - (\ref{Eq24})\\
$\quad h=\sum\nolimits_{i = 1}^{ + \infty } {{\big( {\begin{smallmatrix}
\alpha(x)-1\\
i-1
\end{smallmatrix}} \big)}\frac{{{f^{(i)}}({x_{k}})}}{{\Gamma (i + 1 - \alpha(x) )}}{{({x_k} - c)}^{i- \alpha(x) }}}$\\
$\quad {x_{k + 1}} = {x_k} - \mu  h$\\
$\bf{end\ for}$\\
\hline
\end{tabular*}
\end{table}
\vspace{-3pt}

\begin{remark}\label{Remark 1}
In this paper, three possible solutions are developed to solve the nonconvergence problem of fractional order gradient method. The first method is benefit from the long memory characteristics of fractional derivative, and then the short memory principle is adopted here. The second method is proposed from the infinite series representation of fractional derivative and keep only the first order term. The third method is to change the order with the loss function at each step. What is more, the three ideas can be used in any combination. This work can guarantee that when the considered algorithm is convergent, it will converge to the real extreme point, which will make the fractional order gradient method potential, potent and practical.
\end{remark}
\begin{remark}\label{Remark 2}
In general, it is difficult or even impossible to obtain the analytic form of fractional derivative for arbitrary functions. With regard to this work, the equivalent representation in formulas (\ref{Eq3})-(\ref{Eq4}) plays a critical role. Furthermore, the following formulas (\ref{Eq30})-(\ref{Eq31}) could also be equally adopted.
\begin{equation}\label{Eq30}
{\textstyle{}_{c}^{{\rm{RL}}}\mathscr{D}_x^\alpha f(x) = \sum\nolimits_{i = 0}^{ + \infty } \frac{{{f^{(i)}}(c)}}{{\Gamma (i + 1 - \alpha )}}{{(x - c)}^{i - \alpha }},}
\end{equation}
\begin{equation}\label{Eq31}
{\textstyle{}_{c}^{{\rm{C}}}\mathscr{D}_x^\alpha f(x) = \sum\nolimits_{i = n}^{+\infty}{\frac{{{f^{(i)}}({c})}}{{\mathrm{\Gamma} (i + 1 - \alpha )}}} {(x - c)^{i - \alpha }}.}
\end{equation}

Note that all of these equivalent descriptions are given with infinite terms. To avoid the implementation difficulties brought by the infinite series, the approximate finite sum may be alternative. Additionally, if relation between $f\left(\cdot\right)$ and $x$ is known, fractional order tracking differentiator can be adopted here to calculate ${ {{}_{{x_{k - K}}}^{\hspace{9pt}\rm{C}}{\mathscr D}_x^\alpha f\left( x \right)} |_{x = {x_k}}}$ online, i.e., ${}_{{x_{k - K}}}^{\hspace{9pt}\rm{C}}{\mathscr D}_x^\alpha f\left( x \right){|_{x = {x_k}}} = {}_c^{\rm{C}}{\mathscr D}_t^\alpha f(t){|_{t = {x_k}}} - {}_c^{\rm{C}}{\mathscr D}_t^\alpha f(t){|_{t = {x_{k - K}}}}$ where $c<\mathop {\min }\limits_k {x_k}$.
\end{remark}
\begin{remark}\label{Remark 3}
About the possible solutions to solve the convergence problem, some points are provided here.
\begin{enumerate}[i)]
  \item In Algorithm 1, $0<\alpha<1$ is generally selected for (\ref{Eq11}), when ${{({x_k} - {x_{k - K}})}^{i - \alpha }}$ is used. When it is selectively replaced by ${{|{x_k} - {x_{k - K}}|}^{i - \alpha }}$ or ${{(|{x_k} - {x_{k - K}}|+\epsilon)}^{i - \alpha }}$, the range of $\alpha$ can extend to $(0,2)$ for (\ref{Eq15}).
  \item In Algorithm 2, combining with the fixed memory principle, the leading term in (\ref{Eq19}), i.e., ${{(|{x_k} - c|+\epsilon)}^{i - \alpha }}$ could be modified by ${{(|{x_k} - x_{k-K}|+\epsilon)}^{i - \alpha }}$ and then the range of its application is enlarged to $0<\alpha<2$.
  \item In Algorithm 3, ${{({x_k} - c)}^{i - \alpha(x) }}$ could also find its substitute as ${{|{x_k} - c|}^{i - \alpha(x) }}$ or ${{(|{x_k} - c|+\epsilon)}^{i - \alpha(x) }}$.
  \item Note that (\ref{Eq20})-(\ref{Eq24}) are not the unique forms of the order and any valid forms are suitable here.
  \item Only Caputo definition is considered in constructing the solutions, while the Riemann--Liouville case can still be handled similarly.
  \item Although this paper only focuses on the scalar fractional gradient method ($x\in\mathbb{R}$), the multivariate case can be also established with the similar treatment ($x\in\mathbb{R}^m$, $m>1$).
\end{enumerate}
\end{remark}
\begin{remark}\label{Remark 4}
This paper mainly discusses the possibility on true convergence of fractional order gradient methods. 
\begin{enumerate}[i)]
  \item Compared with \cite{Pu:2015TNNLS}, this work points out the specific reasons for the nonconvergence of fractional order gradient method.
  \item Compared with \cite{Zahoor:2009EJSR}, this work considers a more general target function and proposes three effective solutions to ensure the convergence point is the real extreme point.
  \item Algorithm 1 is actually the generalization of the method in \cite{Chen:2017AMC} and when $K=1$, Algorithm 1 degenerates into the one in \cite{Chen:2017AMC}.
  \item Algorithm 2 and Algorithm 3 are newly developed in this paper. The main ideas implied in Algorithm 1-Algorithm 3 are not contradictory and they can be combined as required.
  \item The work also provides online computation methods for fractional derivative, i.e., series representation and tracking differentiator.
\end{enumerate}
\end{remark}

\section{Simulation Study}\label{Section 4}
In this section, several examples are provided to explicitly demonstrate the validity of the proposed solutions. Examples \ref{Example 1}-\ref{Example 3} aim at testifying the convergence design with the given univariate target function in Section 2.2. Examples \ref{Example 4}-\ref{Example 5} consider a convex and non-convex target functions respectively. Examples \ref{Example 6} gives the application regarding to LMS algorithm. Examples \ref{Example 7} compares the developed method with the classical gradient method.

\begin{example}\label{Example 1}
\textbf{Univariate target function with Algorithm 1}

Recalling $f(x)=(x-5)^2$ and setting the fractional order $\alpha=0.7$, the learning rate $\mu=0.3$, the initial \textcolor[rgb]{0,0,1}{search} point $x_0=1$, then the results using Algorithm 1 with different $K$ are given in the left figure of Fig. \ref{Fig 3}. When $K=1$, $\mu$ varies from $0.1$ to $0.5$ and the simulation with Algorithm 1 is conducted once again.
\vspace{-8pt}
\begin{figure}[!htbp]
\centering
\includegraphics[width=0.5\textwidth]{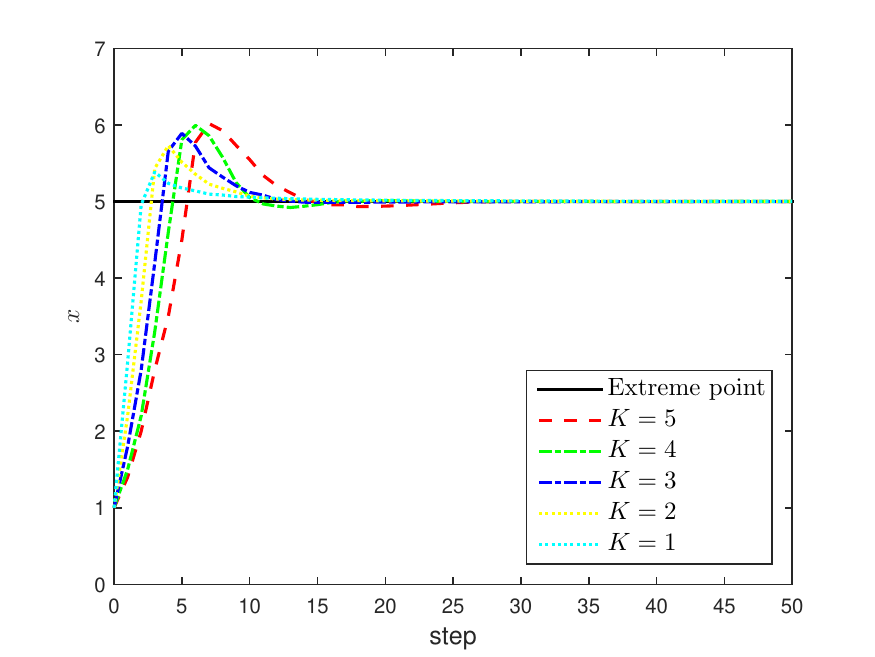}\includegraphics[width=0.5\textwidth]{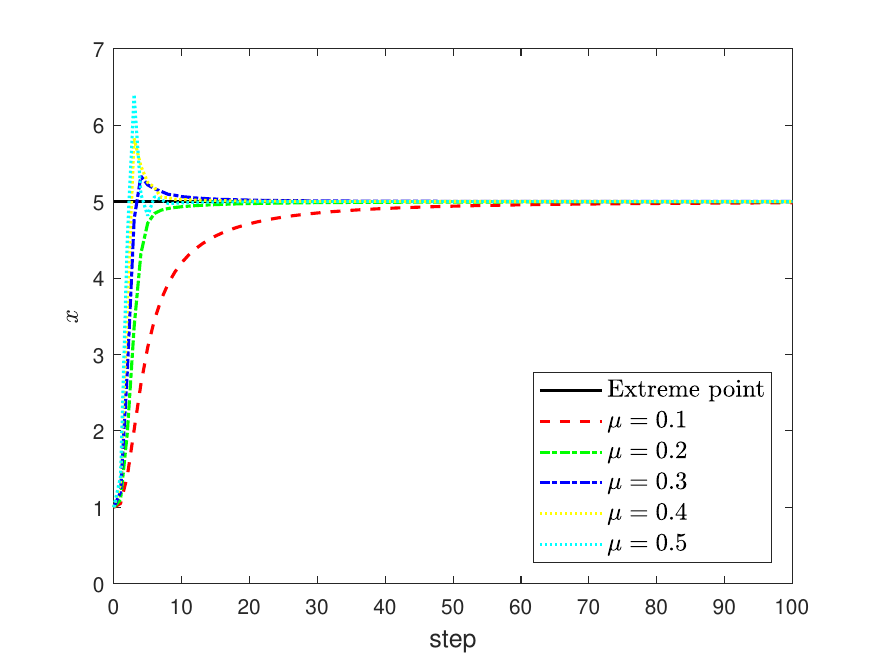}
\centering
\caption{Algorithm 1 (left: performance with different $K$, right: performance with different $\mu$).}\label{Fig 3}
\end{figure}
\vspace{-4pt}

It is clearly seen from the left figure of Fig. \ref{Eq3} that for any case, the expected convergence can reach within $25$ steps. Additionally, the speed of convergence becomes more and more rapid as $K$ decreases. The right one of Fig. \ref{Eq3} indicates that with the increase of the learning rate, the convergence gets faster and faster. In addition, the overshoot emerges gradually.
\end{example}
\begin{example}\label{Example 2}
\textbf{Univariate target function with Algorithm 2}

Let us continue to consider the target function $f(x)=(x-5)^2$.
Setting $c=0$, $\mu=0.15$, $\epsilon=0$, $\alpha=0.7$ and $x_0=1.0,1.5,\cdots,6.0$, Algorithm 2 is adopted to search the minimum value point.
\vspace{-8pt}
\begin{figure}[!htbp]
\centering
\includegraphics[width=0.5\textwidth]{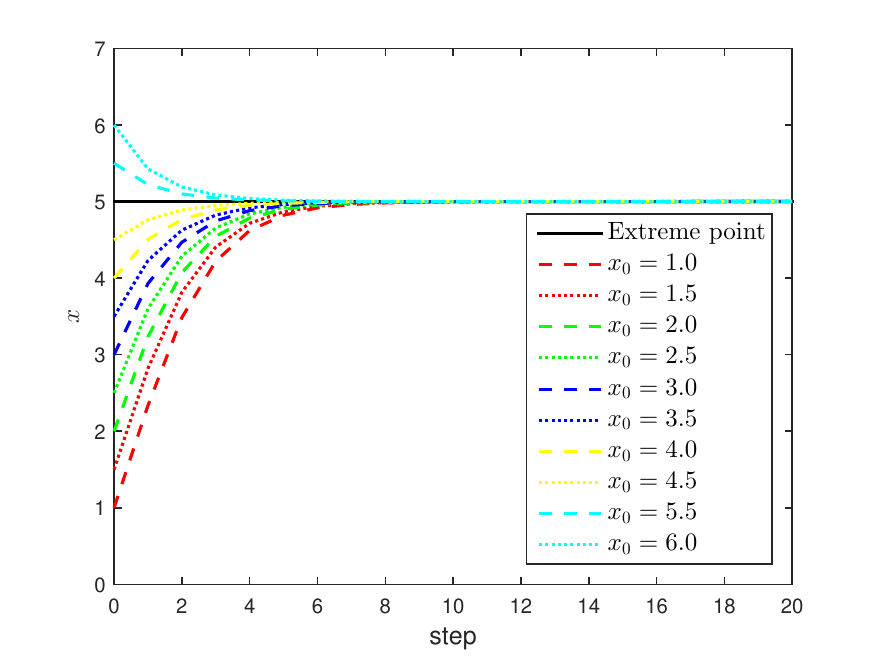}\includegraphics[width=0.5\textwidth]{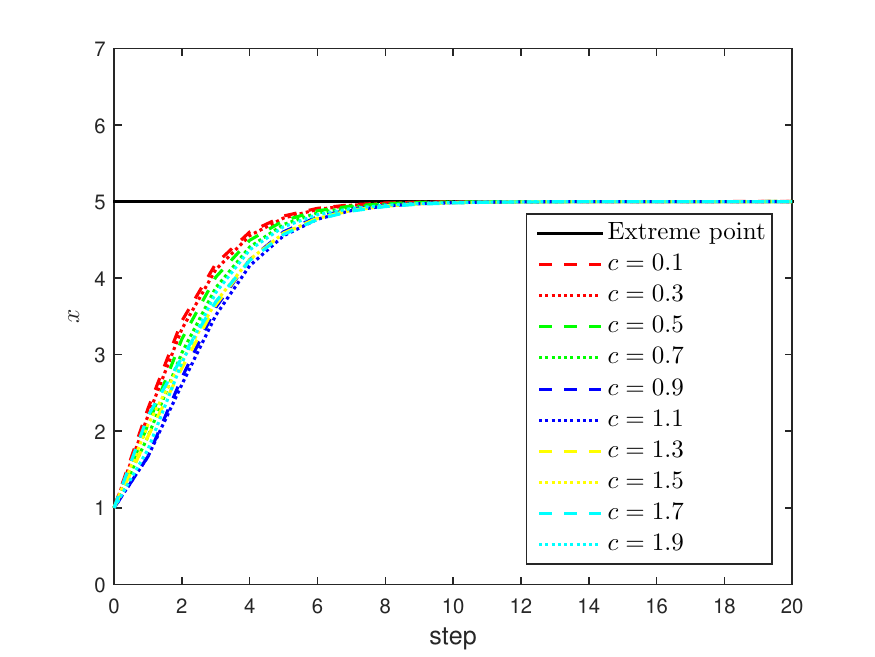}
\centering
\caption{Algorithm 2 (left: performance with different $x_0$, right: performance with different $c$).}\label{Fig 4}
\end{figure}
\vspace{-4pt}

It can be clearly observed from the left one of Fig. \ref{Fig 4} that the proposed method is effective and the algorithm with $x_0$ converges simultaneously. The convergence of Algorithm 1 is robust to the initial search point $x_0$. Similarly, provided $x_0=1$ and $c=0.1,0.3,\cdots,1.9$, the related simulation is performed and the results are shown in the right figure of Fig. \ref{Fig 4}. This picture suggests that the algorithm with different lower integral terminal can truly converge. When $c$ is far from $x_0$, the convergence will accelerate.
\end{example}

\begin{example}\label{Example 3}
\textbf{Univariate target function with Algorithm 3}

Look into the target function $f(x)=(x-5)^2$ once again. Five cases in (\ref{Eq27}) are considered and other parameters are set as $c=0$, $\mu=0.15$ and $x_0=1$. The corresponding results are depicted in the left one of Fig. \ref{Fig 5}.
\vspace{-8pt}
\begin{figure}[!htbp]
\centering
\includegraphics[width=0.5\textwidth]{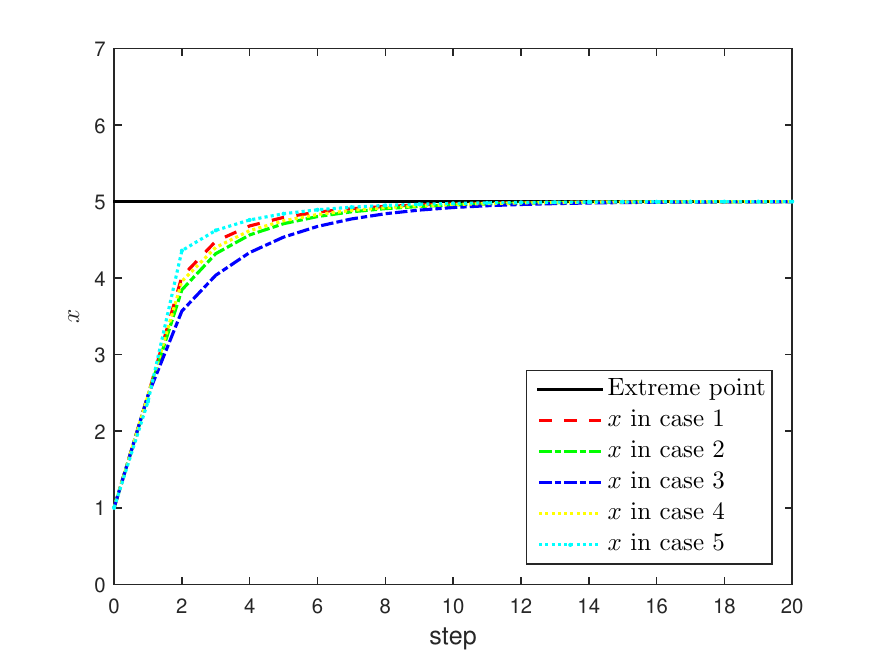}\includegraphics[width=0.5\textwidth]{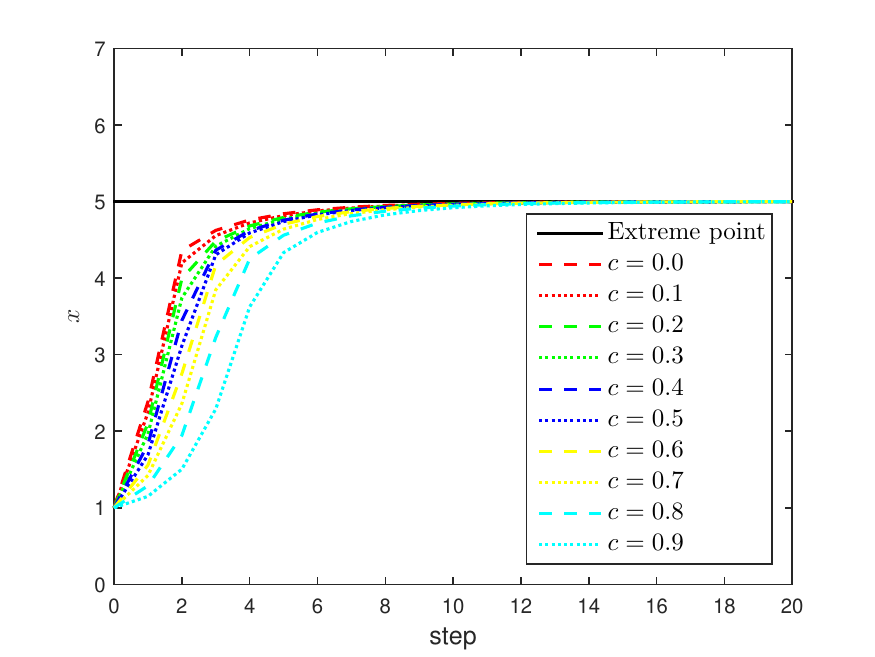}
\centering
\caption{Algorithm 3 (left: performance with different $\alpha(x)$, right: performance with different $c$).}\label{Fig 5}
\end{figure}
\vspace{-4pt}

It is shown that all the designed orders could achieve the convergence as expected while there is still a valuable work to design suitable order for better performance. In the previous simulation, the lower integral terminal $c$ is randomly selected and different from $x_0$. To test the influence of $c$, a series of values $0.0,0.1,\cdots,0.9$ are configured for $c$ in case 5 and the results are shown in the right one of Fig. \ref{Fig 5}. It illustrates that when $c$ increases, the search process gets slower since $|x_0-c|$ decreases accordingly. However, all of them are convergent as expected. The lower integral terminal does not affect the convergence of Algorithm 3.
\end{example}

\begin{example}\label{Example 4}
\textbf{Convex multivariable target function}

Consider the target function proposed in \cite{Pu:2015TNNLS}, i.e., $f(x,y)=2(x-5)^2+3(y-6)^2+10$. The real extreme point of $f(x,y)$ is $x=5,y=6$ and the extreme value is nonzero. Setting $c=0$, $x_0=1$, $y_0=1$, $\mu=0.05$, $\epsilon=0$, $\alpha=0.7$, $K=3$ and $\beta=0.005$, then the results using three proposed methods are given in Fig. \ref{Fig 6} and Fig. \ref{Fig 7}.

\begin{figure}[!htbp]
\centering
\includegraphics[width=0.5\textwidth]{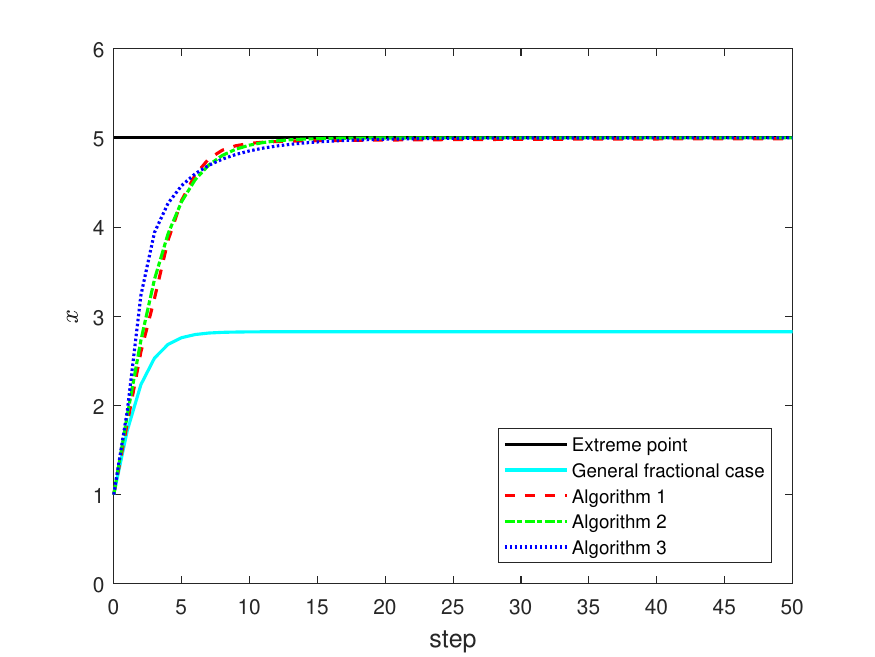}\includegraphics[width=0.5\textwidth]{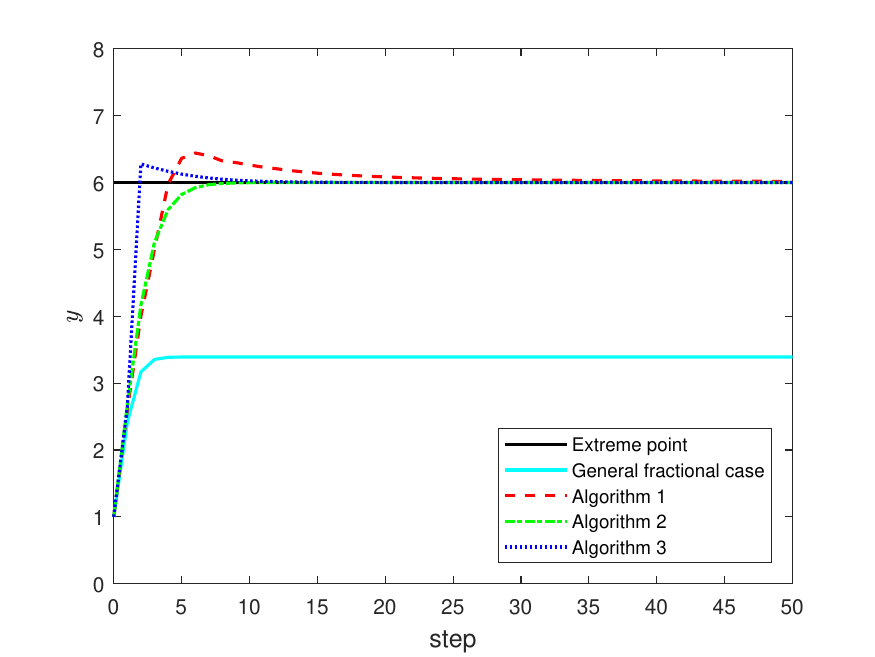}
\centering
\caption{Search results for Example \ref{Example 4} (left: variation of $x$, right: variation of $y$).}\label{Fig 6}
\end{figure}
\begin{figure}[!htbp]
\centering
\includegraphics[width=0.5\textwidth]{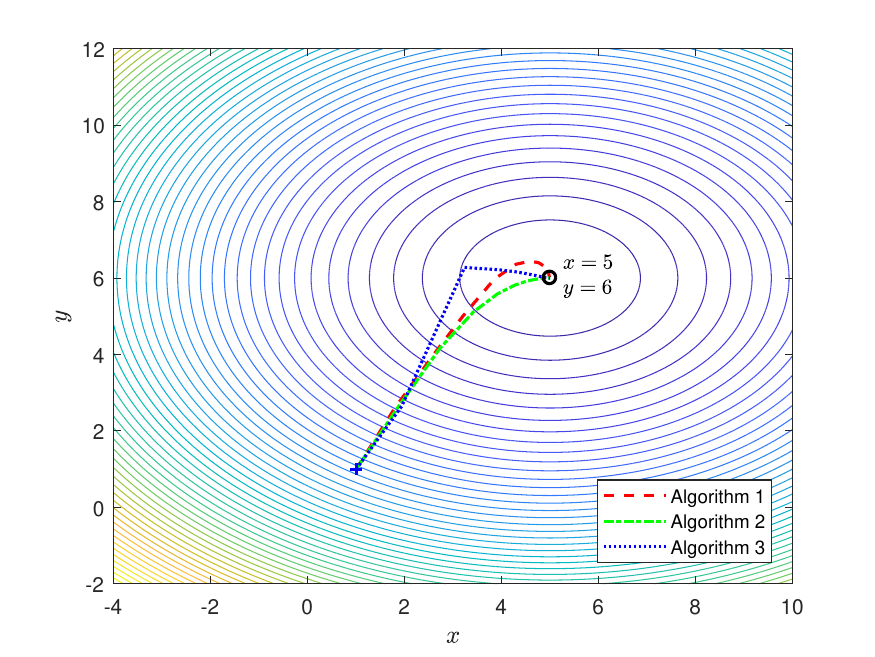}
\centering
\caption{Contour plot for Example \ref{Example 4}.}\label{Fig 7}
\end{figure}

As revealed in Fig. \ref{Fig 6}, both the algorithm from \cite{Pu:2015TNNLS} and the proposed algorithms converge in the $x$ and $y$ coordinate directions. The solid cyan line manifests that the convergent point of the existing algorithm from \cite{Pu:2015TNNLS} is not equal to the real extreme point, which just coincides with the claim in \cite{Pu:2015TNNLS}. Luckily, all the three proposed algorithms converge to the real extreme points as expected. To give a more intuitive understanding, the convergence trajectories are displayed in Fig. \ref{Fig 7}. The iterative search process of $y$ plays faster than that of $x$, since the coefficient about $y$ is larger than that $x$ under the same condition. Besides, extra simulation indicates that when the learning rates are set separately as $\mu_x$ and $\mu_y$ instead of $\mu$, the convergence on $x$ will speed up and the overshoot on $y$ will disappear with properly selected $\mu_x$, $\mu_y$.

Notably, the variable fractional orders are designed individually, namely,
\begin{equation}\label{Eq32}
{\textstyle
\left\{ \begin{array}{l}
\alpha (x) = 1 - \tanh ( {\beta J_x(x,y)} ),\\
\alpha (y) = 1 - \tanh ( {\beta J_y(x,y)} ),
\end{array} \right.}
\end{equation}
where $J_x(x,y) = {\big| {\frac{\partial }{{\partial x}}f\left( {x,y} \right)} \big|^2}$ and $J_y(x,y) = {\big| {\frac{\partial }{{\partial y}}f\left( {x,y} \right)} \big|^2}$. Actually, the order can also be designed uniformly
\begin{equation}\label{Eq33}
{\textstyle
\alpha (x,y) = 1 - \tanh ( {\beta J(x,y)} ),}
\end{equation}
where $J(x,y) = {\big| {\frac{\partial }{{\partial x}}f\left( {x,y} \right)} \big|^2} + \gamma {\big| {\frac{\partial }{{\partial y}}f\left( {x,y} \right)} \big|^2}$ and $\gamma>0$ is the weighting factor.
\end{example}

\begin{example}\label{Example 5}
\textbf{Non-convex multivariable target function}

Consider the famous Rosenbrock function \cite{Rosenbrock:1960CJ}, i.e., $f(x,y)=(1-x)^2+100(y-x^2)^2$. The real extreme point of $f(x,y)$ is $x=1,y=1$. Choose the parameters as $c=0$, $x_0=-0.2$, $y_0=-0.2$, $\epsilon=0$, $\alpha=0.7$, $K=2$ and $\beta=0.01$. Because the minimum value of the target function is $0$, the formula (\ref{Eq24}) is adopted to calculate a common $\alpha(x,y)$ with $J(x,y)=f(x,y)$. The learning rates are selected separately for the three algorithms, i.e., $\mu=0.0182$, $\mu=0.0018$ and $\mu=0.002$. On this basis, three proposed methods are implemented numerically and the simulation curves are recorded in Fig. \ref{Fig 8} and Fig. \ref{Fig 9}.

\begin{figure}[!htbp]
\centering
\includegraphics[width=0.5\textwidth]{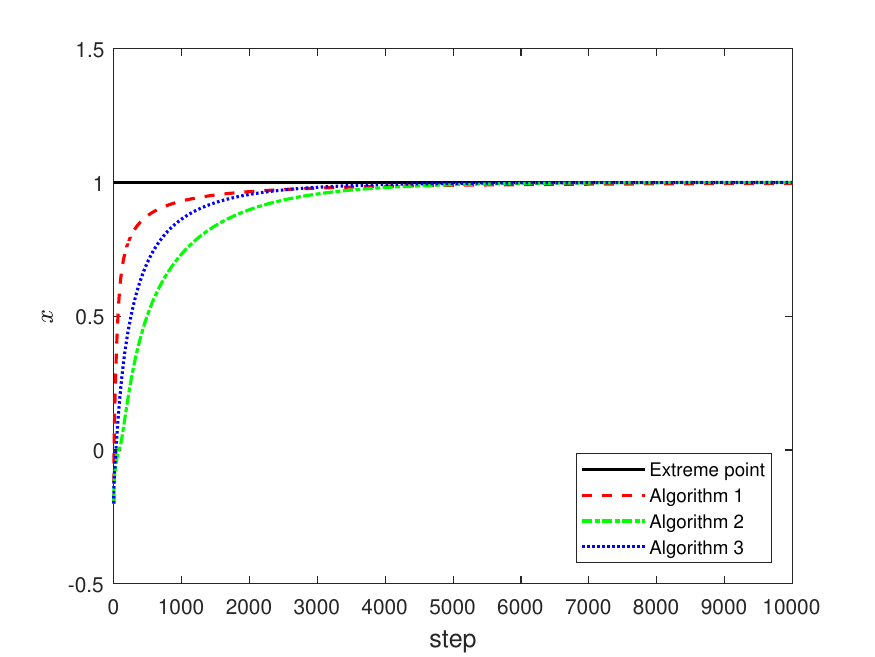}\includegraphics[width=0.5\textwidth]{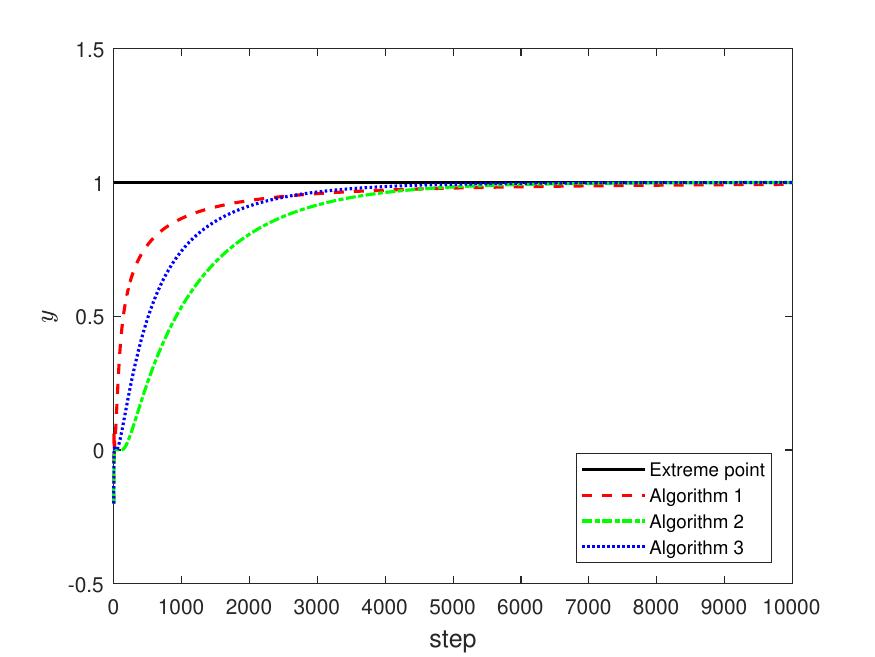}
\centering
\caption{Search results for Example \ref{Example 5} (left: variation of $x$, right: variation of $y$).}\label{Fig 8}
\end{figure}
\begin{figure}[!htbp]
\centering
\includegraphics[width=0.5\textwidth]{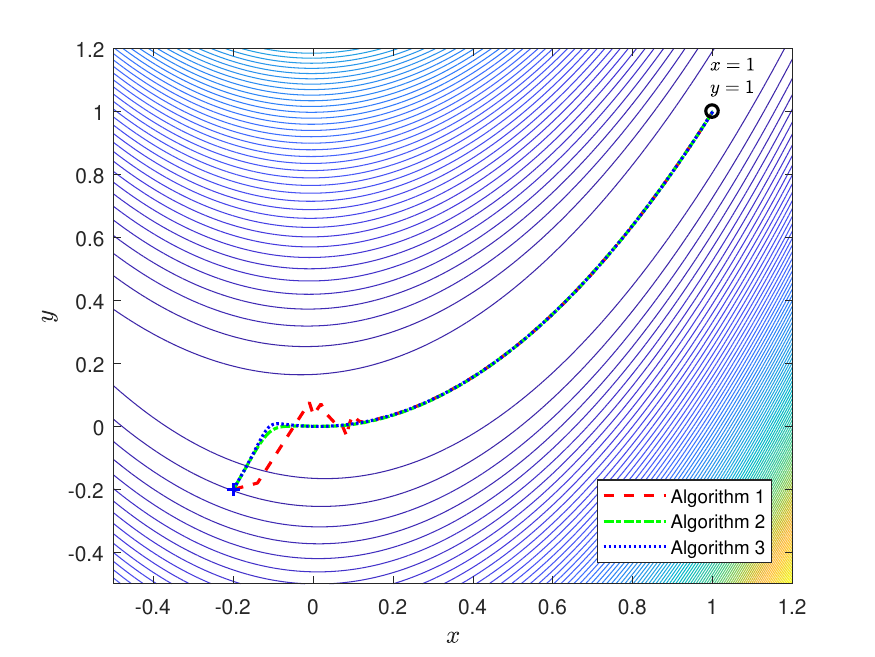}
\centering
\caption{Contour plot for Example \ref{Example 5}.}\label{Fig 9}
\end{figure}

Rosenbrock function is a non-convex benchmark function, which is often used as a performance test problem for optimization algorithm \cite{Rosenbrock:1960CJ}. The extreme point of Rosenbrock function is in a long, narrow and parabolic valley, which is difficult to reach. Fig. \ref{Fig 8} gives variation of $x$ and $y$ of Rosenbrock case, respectively. The general fractional case is not given in the figures because it also converges to a point different from the real extreme point. The three proposed methods are given in figures, and it turns out that Algorithm 2 and Algorithm 3 converge to the extreme point at around $6000$th iteration. Algorithm 1 could approximate the extreme point while small deviations still exist. More specially, defining the error as
\begin{equation}\label{Eq34}
{\textstyle
{\bf e}_k =[x_k,y_k]-[1,1],}
\end{equation}
then it can be calculated that ${\bf e}_{5000}=[0.0105,0.0208]$ and ${\bf e}_{10000}=[0.0035,0.007]$. Though Algorithm 1 performs bad in later period, it is surely ahead of the other algorithms within $5000$th iteration and it is the first one entered the $2\%$ error band. According to Fig. \ref{Fig 9}, what can be seen is that all the curves get together as $y=x^2$ when they are away from the starting point soon. Although Algorithm 1 does not converge to the extreme point in the previous two figures, it converges in the right direction. Therefore, it is plain that the red dotted line in Fig. \ref{Fig 9} would eventually reach to the extreme point $(1,1)$.
\end{example}
\begin{example}\label{Example 6}
\textbf{The application in LMS algorithm}

Consider a three order transverse filtering issue shown in Fig. \ref{Fig 10}. The optimal tap weight is ${{w}}=[w_1,w_2,w_3]=[2,-3,1]$. The known input $u$ and unknown noise $v$ are given in the left one of Fig. \ref{Fig 11}. Select the parameters $c=0$, ${{\hat w}}(0)=[0.1,-0.1,0.1]$, $\mu=0.02$, $\epsilon=0$, $\alpha=0.7$, $K=3$ and $\beta=0.005$. $\alpha(w_i),~i=1,2,3$ are designed according to (\ref{Eq24}) with $J(w)=e^2=(y-\hat y)^2$. Then, all the three proposed methods can be used to estimate parameters of the filter and simulation results are given in the right one of Fig. \ref{Fig 11}.

\begin{figure}[!htbp]
\centering
\includegraphics[width=0.5\textwidth]{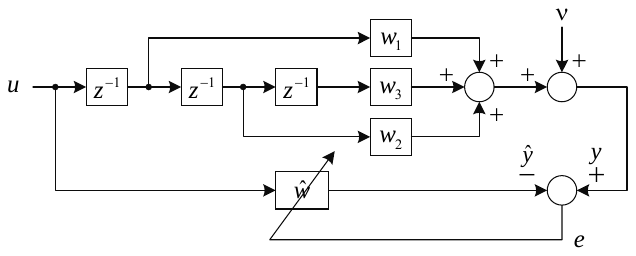}
\centering
\caption{The block diagram of transverse filter.}\label{Fig 10}
\end{figure}
\begin{figure}[!htbp]
\centering
\includegraphics[width=0.5\textwidth]{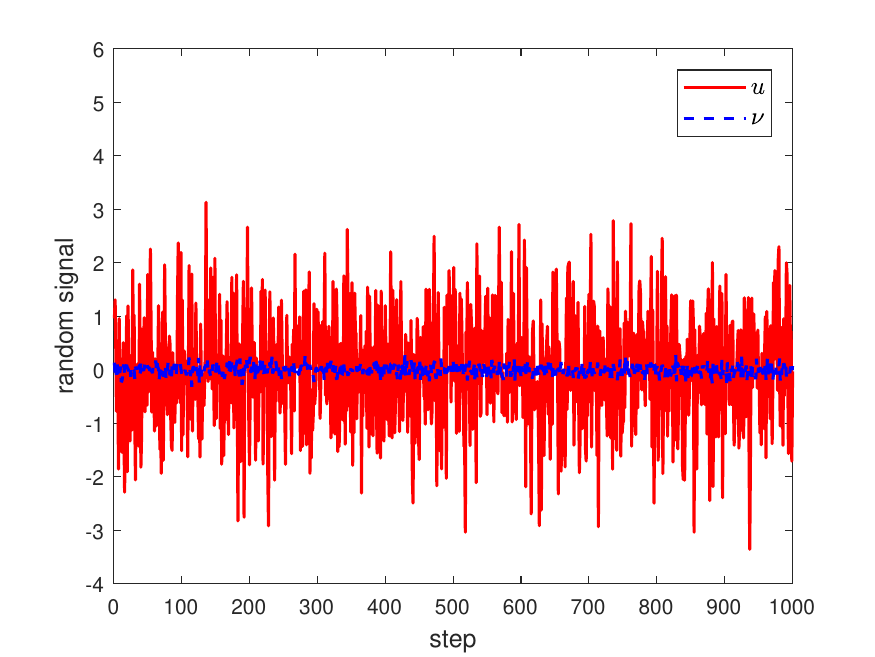}\includegraphics[width=0.5\textwidth]{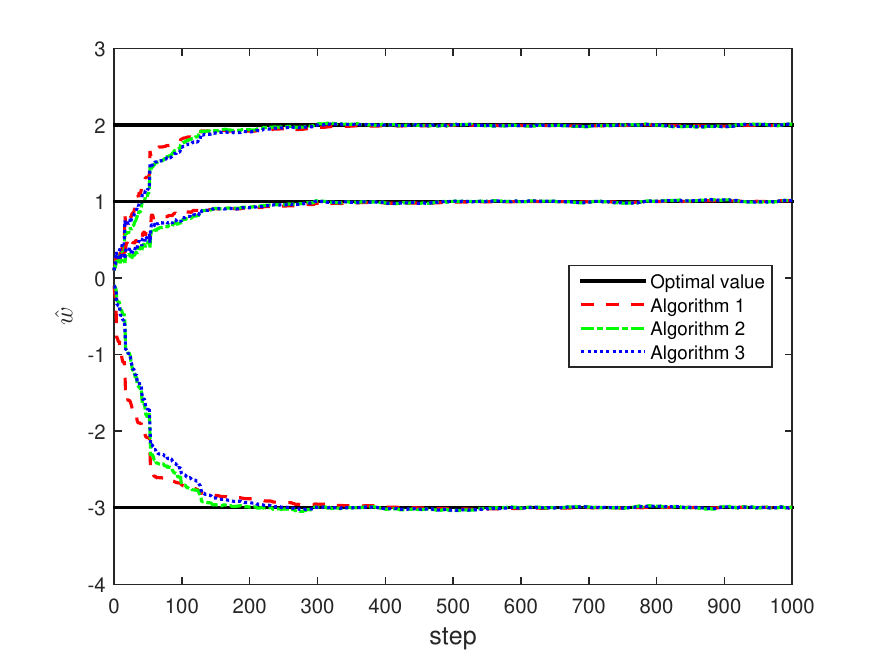}
\centering
\caption{Results for Example \ref{Example 6} (left: known input and unknown noise, right: response of the tap weight ${{\hat w}}(k)$).}\label{Fig 11}
\end{figure}

As can be seen from Fig. \ref{Fig 11}, all of them accomplish the parameter estimation successfully. It is no exaggeration to say that the three methods have considerable convergence speed. Fig. \ref{Fig 11} also demonstrates that the proposed solutions can resolve the convergence problems of fractional order gradient method and the resulting methods are implementable in practical case.
\end{example}
\begin{example}\label{Example 7}
\textbf{The comparison with the classical algorithm}

Herein, we continue to consider the function $f(x)=(x-5)^2$ and check the convergence speed of the classical integer order gradient method in (\ref{Eq5}) and the three developed algorithms. Set the parameters as $\mu=0.15$, $K=1$, $x_0=1$, $c=0$, $\epsilon=0$, and $\beta=0.1$. $x_1=3.4$ is additionally assumed for Algorithm 1 and then the simulation results can be found in Fig. \ref{Fig 12}. From these curves, it can be concluded that our methods overperform the classical method. Though Algorithm 1 plays a little bad result in the steady precision, it gives a rapid response. Notably, with the given conditions, Algorithm 2 and Algorithm 3 perform much better than the integer order method. All of these clearly demonstrate the advantages of the proposed methods in convergence speed.

\begin{figure}[!htbp]
\centering
\includegraphics[width=0.33\textwidth]{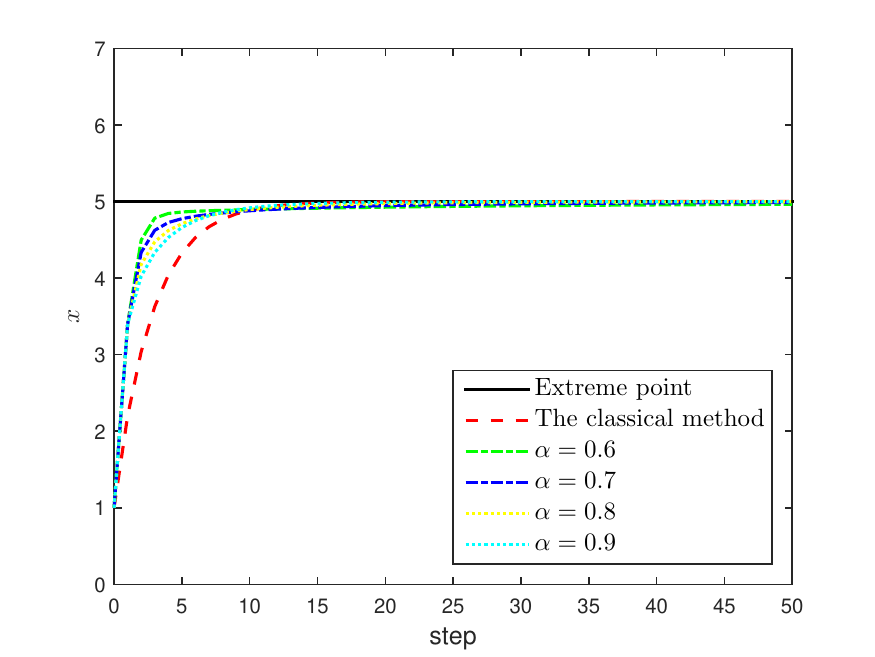}\includegraphics[width=0.33\textwidth]{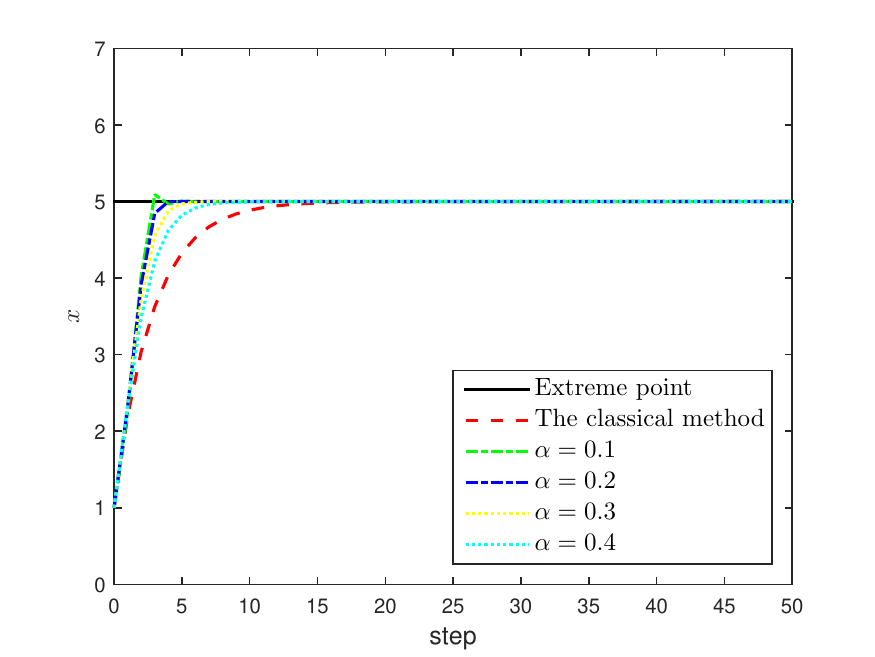}\includegraphics[width=0.33\textwidth]{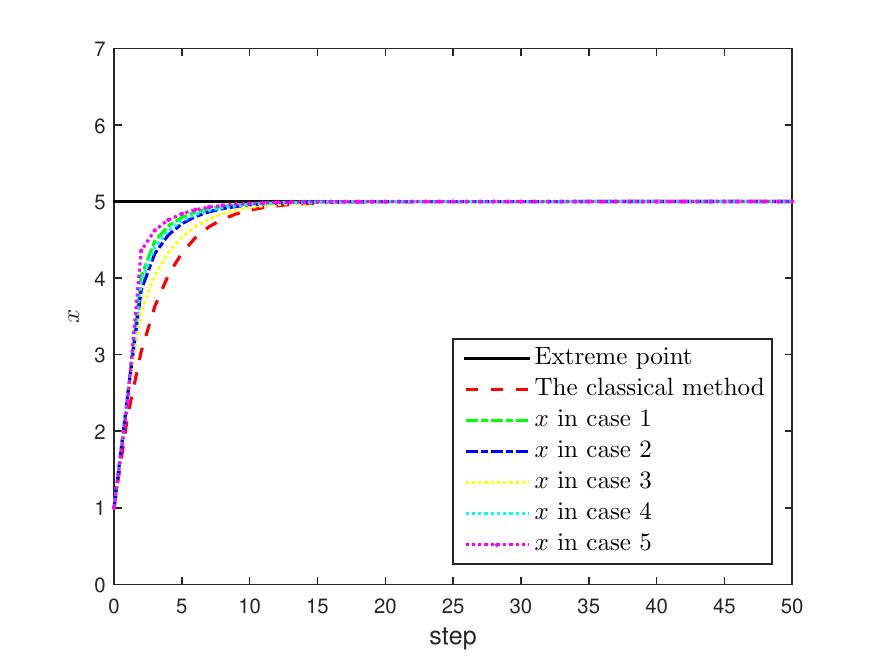}
\centering
\caption{Results for Example \ref{Example 7} with $f(x)=(x-5)^2$ (left: Algorithm 1; middle: Algorithm 2; right: Algorithm 3).}\label{Fig 12}
\end{figure}

\begin{figure}[!htbp]
\centering
\includegraphics[width=0.33\textwidth]{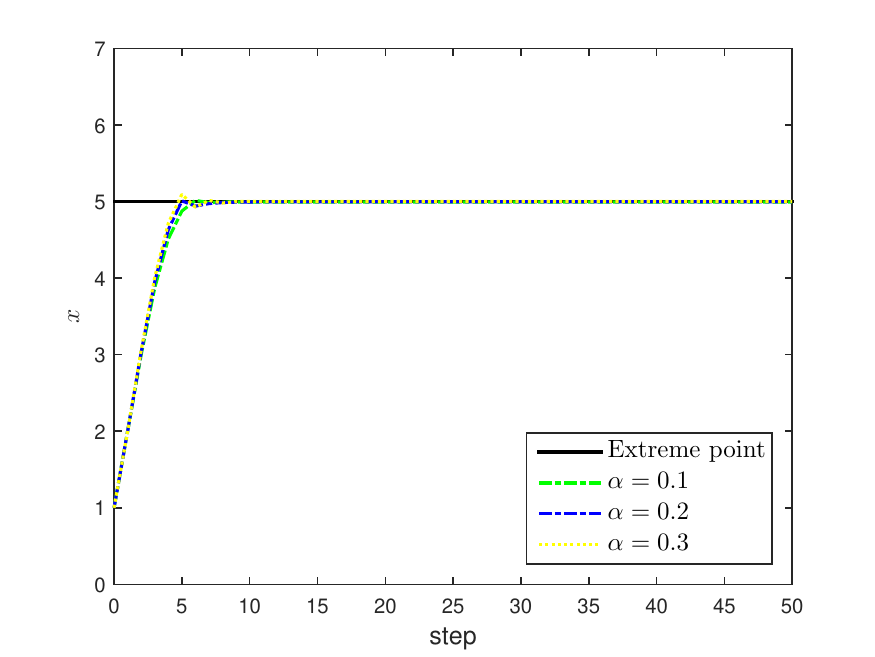}\includegraphics[width=0.33\textwidth]{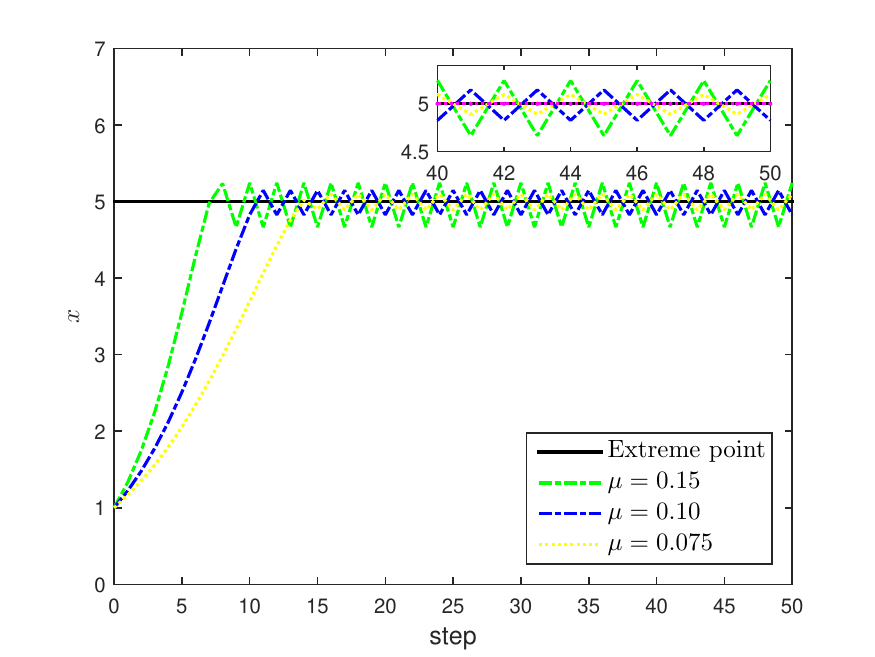}\includegraphics[width=0.33\textwidth]{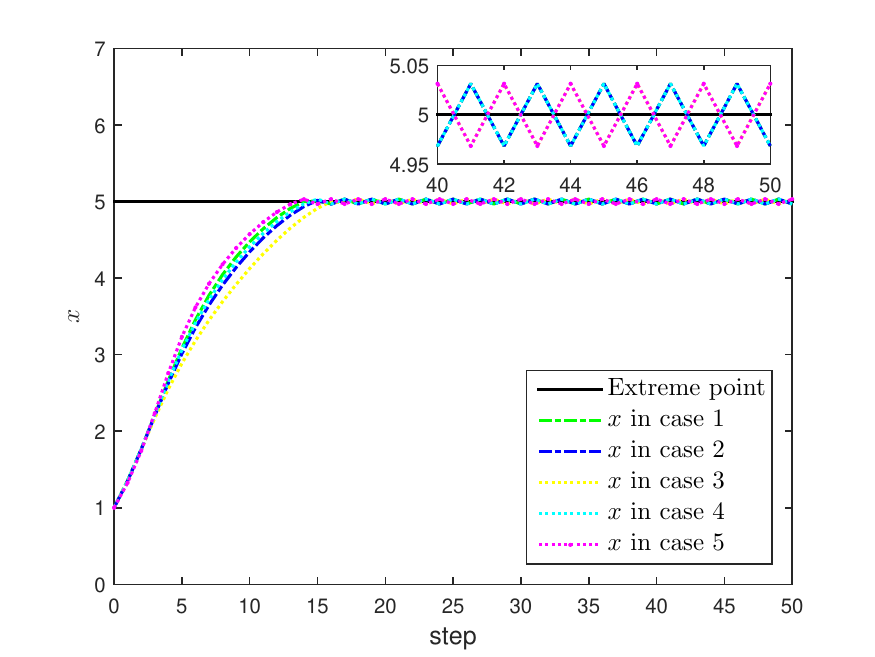}
\centering
\caption{Results for Example \ref{Example 7} with $f(x)=(x-5)^{4/3}$ (left: Algorithm 1; middle: Algorithm 2; right: Algorithm 3).}\label{Fig 13}
\end{figure}

To fully reveal the advantages, we consider the target function $f(x)=(x-5)^{4/3}$, which is not strong convex. By using the method in (\ref{Eq5}), the result is divergent and the real extreme point cannot be found. If we configure the parameters $\mu=0.5$, $x_0=1$, $x_1=2$, $K=1$ for Algorithm 1 and $\alpha=0.1$, $c=0$, $x_0=1$, $\epsilon=0$ for Algorithm 2 and $\mu=0.15$, $x_0=1$, $\beta=0.1$ for Algorithm 3, then the simulation results can be obtained in Fig. \ref{Fig 13}. Several conclusions can be drawn. i) $x_k$ in Algorithm 1 can converge to the real extreme point definitely. ii) Algorithm 2 and Algorithm 3 can converge to a small region around the real extreme point. iii) The chattering phenomenon exists in Algorithm 2 and Algorithm 3. iv) A large learning rate corresponds to a large convergence speed and a large amplitude of vibration. v) All the five cases in (\ref{Eq20})-(\ref{Eq24}) are available for the special function $f(x)=(x-5)^{4/3}$. vi) The variable learning rate method might be useful to solving the chattering problem.
\end{example}
\section{Conclusions}\label{Section 5}
In this paper, the convergence problem of fractional order gradient method has been tentatively investigated. It turns out that general fractional gradient method cannot converge to the real extreme point. By exploiting the natural properties of fractional derivative, three individual solutions are proposed in detail, including the fixed memory step, the higher order truncation, and the variable fractional order. Both theoretical analysis and simulation study indicate that all the designed methods can achieve the true convergence quickly. It is believed that this work is beneficial for solving the pertinent optimization problems with fractional order methods. In future, in-depth studies could be undertaken in these promising directions.
\begin{enumerate}[i)]
  \item Design and analyze new convergence design solutions.
  \item Extend the results to fractional Lipschitz condition.
  \item Consider the nonsmooth or nonconvex target function.
  \item Analyze and design the gradient algorithms in a new way \cite{Jordan:2018ICM,Muehlebach:2019arXiv}.
  \item Combine the developed methods with control systems \cite{Wu:2017SMCA,Sun:2018Auto,Su:2018Auto,Zhang:2018TC,Zhang:2019FSS}.
\end{enumerate}

%
\section*{Acknowledgements}
The authors would like to express their gratitude to the reviewers for their insightful comments and suggestions which greatly improved this work. The work described in this paper was supported by the National Natural Science Foundation of China (61601431, 61973291) and the fund of China Scholarship Council (201806345002).

%

\phantomsection
\addcontentsline{toc}{section}{References}
\section*{References}
\bibliographystyle{model1-num-names}
\bibliography{database}

\end{document}